# GPS scintillation and irregularities at the front of an ionization tongue in the nightside polar ionosphere

Christer van der Meeren[1], Kjellmar Oksavik[1,2], Dag Lorentzen[2,3], Jøran Idar Moen[2,4], and Vincenzo Romano[5]

[1]Birkeland Centre for Space Science, Department of Physics and Technology, University of Bergen, Bergen, Norway, [2]University Centre in Svalbard, Longyearbyen, Norway, [3]Birkeland Centre for Space Science, University Centre in Svalbard, Longyearbyen, Norway, [4]Department of Physics, University of Oslo, Oslo, Norway, [5]Istituto Nazionale di Geofisica e Vulcanologia, Rome, Italy

**Abstract** In this paper we study a tongue of ionization (TOI) on 31 October 2011 which stretched across the polar cap from the Canadian dayside sector to Svalbard in the nightside ionosphere. The TOI front arrived over Svalbard around 1930 UT. We have investigated GPS scintillation and irregularities in relation to this TOI front. This is the first study presenting such detailed multi-instrument data of scintillation and irregularities in relation to a TOI front. Combining data from an all-sky imager, the European Incoherent Scatter Svalbard Radar, the Super Dual Auroral Radar Network Hankasalmi radar, and three GPS scintillation and total electron content (TEC) monitors in Longyearbyen and Ny-Ålesund, we observe bursts of phase scintillation and no amplitude scintillation in relation to the leading gradient of the TOI. Spectrograms of 50 Hz phase measurements show highly localized and variable structuring of the TOI leading gradient, with no structuring or scintillation within the TOI or ahead of the TOI.

## 1. Introduction

During active geomagnetic conditions, especially southward oriented interplanetary magnetic field (IMF), high-density plasma may be convected from a solar-ionized high-density plasma reservoir in the dayside ionosphere, through the cusp region, and across the polar cap to the nightside auroral oval [*Dungey*, 1961; *Weber et al.*, 1984; *Foster and Doupnik*, 1984; *Buchau et al.*, 1985; *Foster*, 1993; *Foster et al.*, 2005; *Moen et al.*, 2008; *Cousins and Shepherd*, 2010; *Oksavik et al.*, 2010; *Zhang et al.*, 2013a; *Nishimura et al.*, 2014]. This plasma may be segmented in the cusp region into discrete 100–1000 km islands of high-density plasma called *F* region polar cap patches. Several possible patch creation mechanisms have been suggested [e.g., *Carlson*, 2012, and references therein], among them is transient magnetopause reconnection, which has been proposed as the dominant segmentation mechanism [*Lockwood and Carlson*, 1992; *Carlson et al.*, 2002, 2004, 2006; *Lockwood et al.*, 2005; *Moen et al.*, 2006; *Lorentzen et al.*, 2010; *Zhang et al.*, 2013b]. In the absence of segmentation, the plasma may instead form a continuous tongue of ionization (TOI) convecting and extending across the polar cap [*Sato*, 1959; *Knudsen*, 1974; *Foster et al.*, 2005]. Areas of enhanced electron densities such as patches and TOIs show up as regions of 630.0 nm airglow emissions from dissociative recombination of $O_2^+$ with *F* region electrons creating $O(^1D)$ [*Wickwar et al.*, 1974; *Hosokawa et al.*, 2011]. These emissions are detectable from ground-based optical instruments [e.g., *Buchau et al.*, 1983; *Weber et al.*, 1984; *Lorentzen et al.*, 2004; *Hosokawa et al.*, 2006; *Moen et al.*, 2007; *Nishimura et al.*, 2014].

Large-scale ionospheric plasma structures such as polar cap patches often contain decameter- to kilometer-scale irregularities, particularly on the edges [*Weber et al.*, 1986; *Basu et al.*, 1990, 1991, 1994, 1998; *Moen et al.*, 2012; *Oksavik et al.*, 2012; *Carlson*, 2012, and references therein]. It has been suggested that the Lockwood-Carlson patch segmentation mechanism may initially structure patches through the Kelvin-Helmholtz instability (KHI), creating seed irregularities which allows the gradient drift instability (GDI) to effectively further structure the patch through its transit across the polar cap [*Carlson et al.*, 2007, 2008]. The GDI works on all gradients except where the gradient is exactly antiparallel to the plasma flow, but it is effective primarily at the trailing edges of density enhancements, and it is considered the dominant instability mode in the polar cap [e.g., *Ossakow and Chaturvedi*, 1979; *Keskinen and Ossakow*, 1983; *Buchau et al.*, 1985; *Tsunoda*, 1988; *Basu et al.*, 1990, 1998; *Coker et al.*, 2004; *Prikryl et al.*, 2011a; *Basu et al.*, 1994; *Gondarenko and Guzdar*, 2004a]. Simulations show that these irregularities may propagate from the trailing edge into the interior of the patch, possibly structuring the whole patch at these scale sizes [*Gondarenko and Guzdar*, 2004b]. Observations have indeed shown that patches may be structured throughout [e.g., *Hosokawa et al.*, 2009].

Irregularities with scale sizes of decameters to kilometers may cause rapid phase and/or amplitude variations in transionospheric signals, such as global navigation satellite system (GNSS) signals [e.g., *Hey et al.*, 1946; *Basu et al.*, 1990, 1998; *Kintner et al.*, 2007]. If the irregularities have sufficiently small scale sizes, the phase and amplitude variations are diffractive in nature and are termed scintillations [e.g., *Mushini et al.*, 2012]. Intense scintillation may cause failure of signal reception due to loss of signal lock [e.g., *Kintner et al.*, 2007], even during solar minimum [*Prikryl et al.*, 2010]. Scintillation is present in both hemispheres [*Prikryl et al.*, 2011b].

Ionospheric scintillations are categorized into amplitude scintillations and phase scintillations. Amplitude scintillations are caused by irregularities with scale sizes of tens of meters to hundreds of meters [e.g., *Kintner et al.*, 2007], more precisely at and below the Fresnel radius, which is approximately 360 m for GPS L1 frequency (1575.42 MHz) and a phase screen altitude of 350 km [e.g., *Forte and Radicella*, 2002]. Amplitude scintillations are normally quantified by the $S_4$ index, which is the standard deviation of the received power normalized by its mean value [*Briggs and Parkin*, 1963]:

$$S_4^2 = \frac{\langle I^2 \rangle - \langle I \rangle^2}{\langle I \rangle^2}$$

Here $I$ is the power, and the brackets indicate the expected value, in practice replaced by temporal averaging [e.g., *Beach*, 2006]. Phase scintillations are caused by irregularities of scale sizes from hundreds of meters to some kilometers [e.g., *Kintner et al.*, 2007] and are normally quantified by the $\sigma_\phi$ index, which is the standard deviation of the detrended carrier phase $\phi$ [*Fremouw et al.*, 1978]:

$$\sigma_\phi^2 = \langle \phi^2 \rangle - \langle \phi \rangle^2$$

The $\sigma_\phi$ index is highly sensitive to refractive larger-scale (lower-frequency) phase variations not pertaining to scintillations [e.g., *Basu et al.*, 1991; *Mushini et al.*, 2012], and careful detrending of the raw phase is needed. Specifically, $\sigma_\phi$ is highly sensitive to the detrending filter cutoff frequency. This will be further detailed in section 5.

Performance of GNSS services at high latitudes is of increasing importance in, e.g., offshore and aviation; and understanding scintillation-inducing irregularities is critical for scintillation modeling, forecasting, and mitigation [e.g., *Moen et al.*, 2013]. Much of the research on scintillation is using a statistical approach [e.g., *Spogli et al.*, 2009; *Alfonsi et al.*, 2011a; *Li et al.*, 2010; *Prikryl et al.*, 2010, 2011a; *Tiwari et al.*, 2012]. Some case studies in the Svalbard region have been performed [*De Franceschi et al.*, 2008; *Coker et al.*, 2004; *Mitchell et al.*, 2005]. These studies focus mainly on patches. While patches are known to be structured, it is less known how the TOI is structured. It is readily conceivable that the irregularity processes may differ: Patches may be initially structured by the KHI, and these irregularities are further developed by the GDI. Continuous tongues will not experience this initial KHI structuring. Furthermore, depending on the degree of large-scale structuring within the tongue, there may not be any clear gradients on which the GDI can work. On the other hand, TOIs are regions of steep density gradients where we might expect strong irregularities to be present.

In order to better understand irregularities in relation to TOIs, we present findings of GPS scintillation in relation to the arriving front of a tongue of ionization in the nightside polar cap over Svalbard, using GPS scintillation and total electron content (TEC) monitors, the European Incoherent Scatter (EISCAT) Svalbard Radar (ESR), an optical all-sky airglow imager, and backscatter from Super Dual Auroral Radar Network (SuperDARN) high-frequency (HF) radars. To our knowledge, this is the first paper presenting such a detailed multi-instrument observation of scintillation in the Svalbard region in relation to a TOI.

## 2. Instrumentation
### 2.1. GPS Receivers
The GPS data in this study come from three NovAtel GSV4004B GPS Ionospheric Scintillation and TEC Monitors. Two receivers are located in Ny-Ålesund (78.9° GLAT, 11.9° GLON, 76.6° Altitude Adjusted Corrected

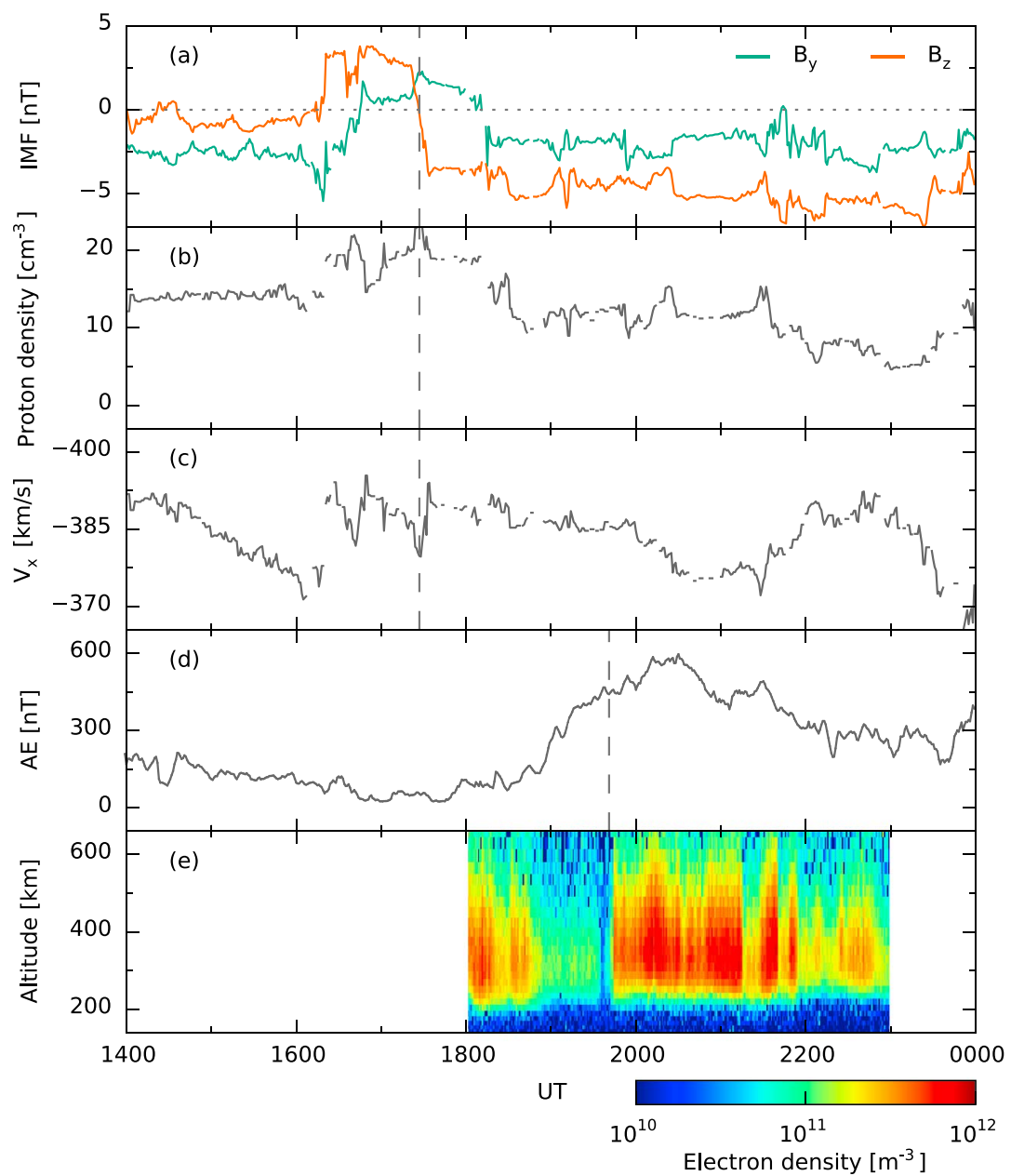

**Figure 1.** Overview of solar wind parameters, substorm activity, and vertical electron density profiles from the EISCAT Svalbard Radar (ESR) on 31 October 2011: (a) IMF $B_y$ in green and $B_z$ in red, (b) solar wind proton density, (c) solar wind speed, (d) the *AE* index, and (e) ESR vertical electron density profiles over Longyearbyen. The vertical line in Figures 1a–1c marks the time of southward turning of the IMF. The vertical line in Figure 1d corresponds to the sharp gradient in Figure 1e and the time shown in Figure 3c (1941 UT).

Geomagnetic Coordinates (AACGM) latitude [*Baker and Wing*, 1989]) and one in Longyearbyen (78.1° GLAT, 16.0° GLON, 75.5° AACGM latitude).

One of the Ny-Ålesund receivers (hereafter NYA) is operated by the University of Oslo (UiO), Norway. Only 60 s data were available from this receiver. The software produces calibrated TEC mapped to vertical TEC (VTEC) and the $S_4$ amplitude scintillation index. The software does not produce the phase scintillation index $\sigma_\phi$ or the rate of TEC (ROT). The real-time software is developed by *Carrano and Groves* [2006], and the data have been postprocessed using WinTEC-P [*Carrano et al.*, 2009], an extension of WinTEC [*Anghel et al.*, 2008].

The other receiver in Ny-Ålesund (hereafter NYA1) and the receiver in Longyearbyen (hereafter LYB0) are operated by the Istituto Nazionale di Geosica e Vulcanologia (INGV, Italy), as part of the Ionospheric Scintillations Arctic Campaign Coordinated Observations (ISACCO) project [*Franceschi et al.*, 2006]. Data are available online by means of the electronic Space Weather upper atmosphere system [*Romano et al.*, 2008, 2013]. These receivers use NovAtel software [*NovAtel*, 2007]. Raw high-resolution data were available from these two receivers. Uncalibrated slant TEC was available at a 1 Hz cadence. The TEC has been tuned to correspond better to the TEC data of the UiO-operated NYA receiver. This is sufficient for our purposes since we do not require absolute TEC measurements. Furthermore, the slant TEC has been mapped to vertical TEC as done by *Alfonsi et al.* [2011b]. ROT was also available at a 0.1 Hz cadence, but due to noisy data in the LYB0

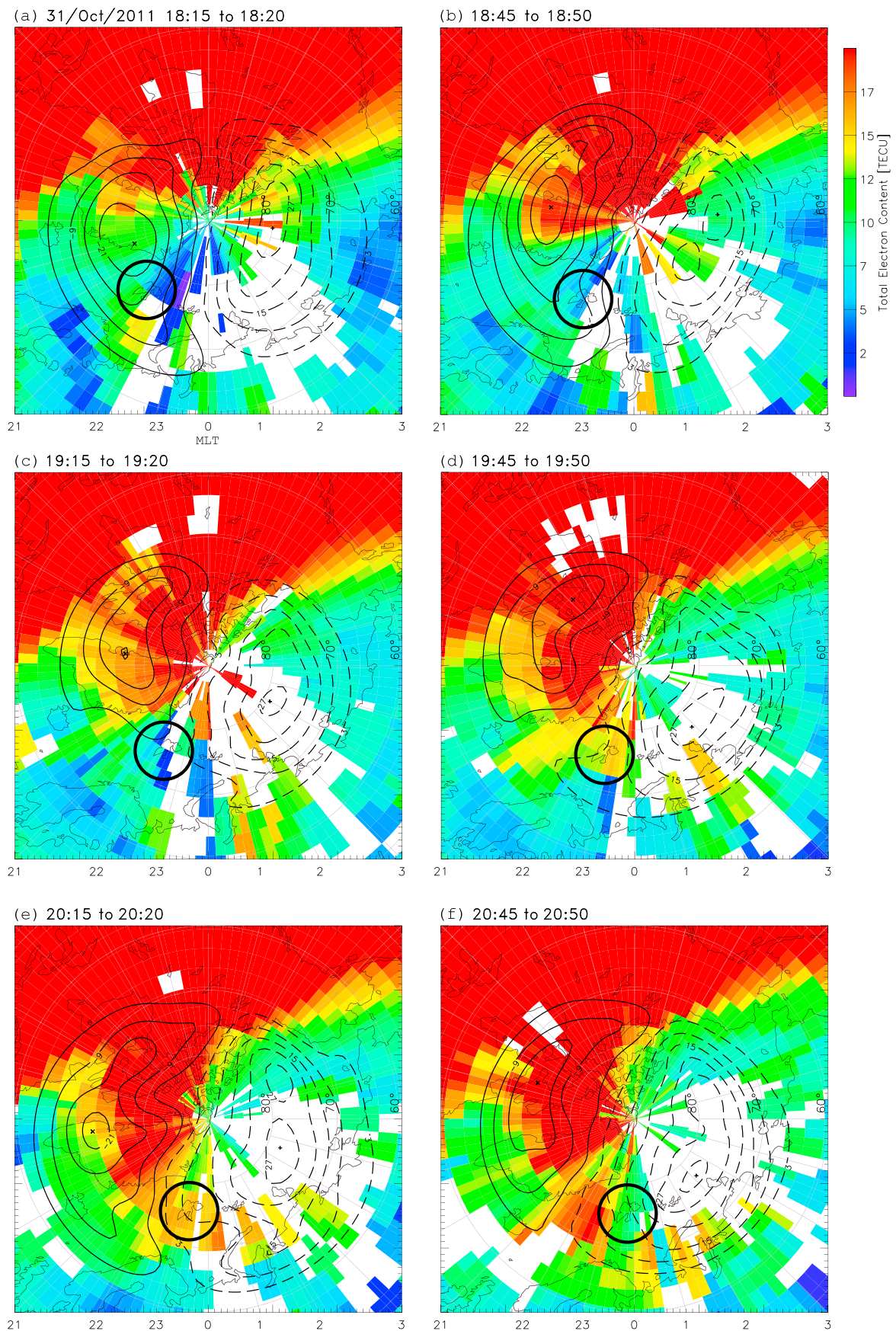

**Figure 2.** (a–f) The progression of the ionization tongue as it follows the dusk cell of the convection pattern across the polar cap. Svalbard is marked by a black circle. Magnetic local time (MLT) is shown on the horizontal axis.

1 Hz ROT, only 60 s ROT is presented for this receiver. ROT is not influenced by satellite-receiver biases. The $S_4$ amplitude scintillation index has been calculated by subtracting the $S_4$ due to ambient noise from the total $S_4$ in a root-sum-square sense according to the user manual [*NovAtel*, 2007]. The 60 s $\sigma_\phi$ phase scintillation index in this study is presented using both a 0.1 Hz and 0.3 Hz detrending filter cutoff frequency. The 0.1 Hz version is calculated by the receiver software in real time, while the 0.3 Hz version is calculated from the 50 Hz raw phase data by us.

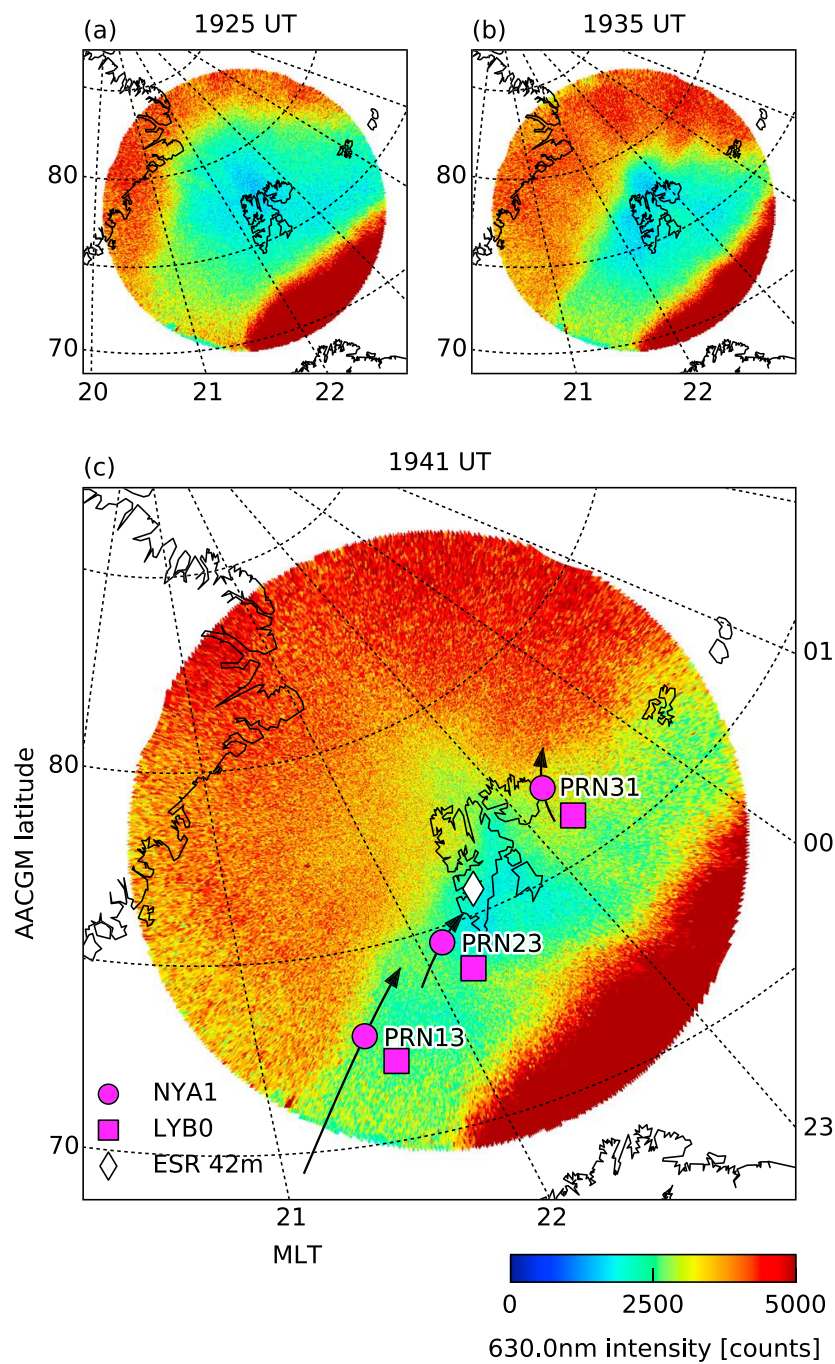

**Figure 3.** All-sky imager (ASI) data showing the propagation of the (a and b) ionization front leading up to its arrival over Longyearbyen at (c) 1941 UT. A geomagnetic grid is shown using Altitude Adjusted Corrected Geomagnetic Coordinates (AACGM) latitude and magnetic local time (MLT). GPS and ESR ionospheric piercing points (IPPs) are shown in Figure 3c. IPPs are shown for both receiver locations (Ny-Ålesund and Longyearbyen) for three different satellites (PRN13, PRN23, and PRN31). Arrows on the GPS IPPs indicate direction and orbit (IPP location) at 1941 UT ±30 min. The saturation in the lower right of the ASI data is due to auroral emissions. At 300 km projection altitude, the diameter of the all-sky projection is ∼1700 km.

For all receivers, some limits were imposed on data validity according to *NovAtel* [2007]. For TEC, ROT, and $S_4$ data, a lock time of at least 60 s has been required. For $\sigma_\phi$ data, a lock time of at least 240 s has been required. In addition to this, we require the elevation to be at least 20° for all data in order to remove clutter associated with multipath signals near the horizon.

### 2.2. Radars

The EISCAT Svalbard Radar (ESR) is a 500 MHz incoherent scatter radar located in Longyearbyen (78.2° GLAT, 16.0° GLON, 75.5° AACGM latitude). The radar consists of two antennae: One 42 m antenna of fixed, field-aligned orientation (81.6° elevation, 182.1° azimuth), and one fully steerable 32 m antenna. In this study, data from the 42 m antenna are presented. We also use data from the 32 m antenna, pointing toward magnetic south at 30° elevation, in the estimation of the plasma drift velocity and for determining the optimal projection altitude. The data have been analyzed using GUISDAP (Grand Unified Incoherent Scatter Design and Analysis Package) [*Lehtinen and Huuskonen*, 1996].

The Super Dual Auroral Radar Network (SuperDARN) is a network of coherent HF scatter radars in both hemispheres measuring backscatter from field-aligned decameter-scale ionospheric irregularities [*Greenwald et al.*, 1995; *Chisham et al.*, 2007]. In this study we present data from the Hankasalmi SuperDARN radar, which has a field of view covering Svalbard.

### 2.3. Optics

The all-sky imager (ASI) used in this study is operated by the University of Oslo and located at Ny-Ålesund. We will present data from the 630.0 nm (red) interference filter in order to study airglow from the recombination of *F* region electrons with molecular oxygen. The intensity of the airglow is proportional to the plasma density at *F* region altitudes [e.g., *Hosokawa et al.*, 2011]. The camera is not calibrated and thus the intensity readings will be given in raw counts.

### 2.4. Solar Wind

Solar wind data are provided by the NASA OMNIWeb service. The data come from the Wind spacecraft, specifically the two instruments Magnetic Field Investigation [*Lepping et al.*, 1995] and Solar Wind Experiment [*Ogilvie et al.*, 1995]. Wind was located at $X = 257\,R_E$, $Y = 32\,R_E$, and $Z = 22\,R_E$ (geomagnetic solar ecliptic coordinates). The data are time shifted to the bow shock by the OMNIWeb service.

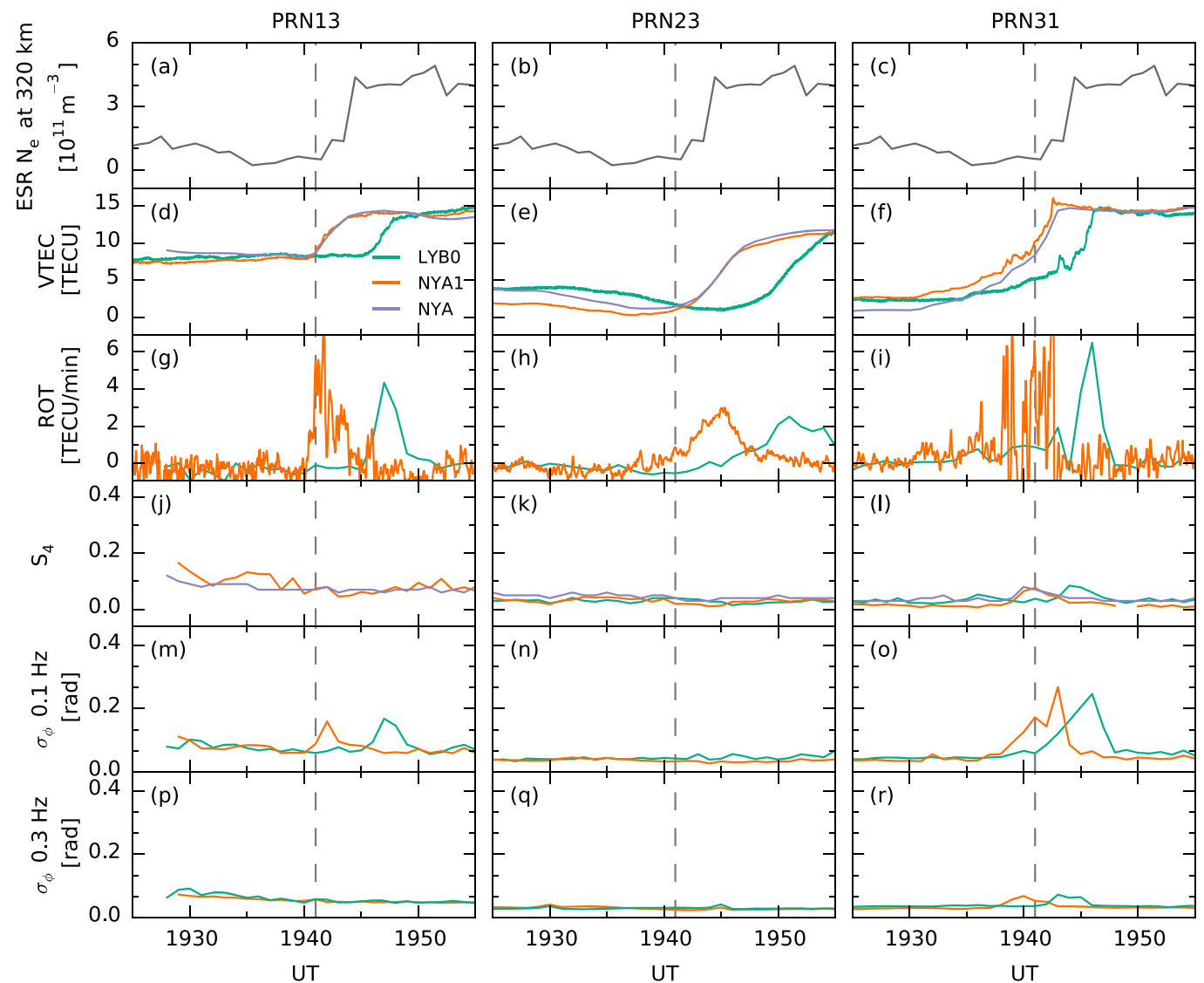

**Figure 4.** GPS data from three satellites (one per column). (a–c) The panels are identical and show the ESR electron density at 320 km altitude. The GPS parameters are (d–f) vertical total electron content (VTEC), (g–i) rate of TEC (ROT), (j–l) the $S_4$ amplitude scintillation index (not available for PRN13 in the LYB0 receiver), and the $\sigma_\phi$ phase scintillation index calculated using a (m–o) 0.1 Hz cutoff frequency and (p–r) a 0.3 Hz cutoff frequency. The vertical dashed lines mark the time shown in Figure 3c. The 0.3 Hz-detrended $\sigma_\phi$ has been calculated by first subtracting a fourth-order polynomial fit from the raw phase, and then filtering the result using a high-pass Butterworth filter with a cutoff frequency of 0.3 Hz. The standard deviation was then computed for each minute of data.

## 3. Geographic Projection of Data

The optimal altitude for the geographic projection of the GPS data was determined by two separate methods. The first method was correlating the TEC values of GPS PRN20 and the electron density ($n_e$) values from the ESR 32 m antenna at a time when the GPS ionospheric piercing point (IPP) closely followed the ESR 32 m beam. The second method was the determination of the altitude of the maximum F region density around the time of interest, based on the assumption that most of the effects on the GPS signal are located in this region. Both methods yielded a result of 320 km for the projection altitude. In order to further test the phase screen model we also verified, using the ESR, that the vertical electron density profile was in fact dominated by F region plasma.

The geographic projection of the ASI data were done using an altitude of 300 km, as recommended by literature [*Hosokawa et al.*, 2011; *Lorentzen et al.*, 2004; *Millward et al.*, 1999]. This projection is sufficient for our purposes because the ASI data are only used as an indicator of enhanced plasma densities in the F region.

## 4. Observations
### 4.1. Overview

A campaign was run at the ESR on 31 October 2011, 18–23 UT. Figure 1 shows an overview of solar wind parameters, substorm activity, and the ESR vertical electron density profiles on this day. The vertical line in Figures 1a–1c marks the time of a southward turning of the IMF, which causes high-density plasma from the dayside to be convected into the polar cap. Figure 1d shows the *AE* index, whose positive bay signature is indicative of a substorm. Figure 1e shows electron density from the ESR 42 m field-aligned antenna. When the front of the TOI passes overhead at 1941 UT, the ESR data show a sudden enhancement of *F* region electron density. This is further enhanced, though not as suddenly, until 2000 UT. The enhanced densities seem to persist until a sudden drop around 2115 UT. Figure 2 shows the progression of the TOI across the polar cap in the antisunward portion of the dusk convection cell. The TOI begins to form on the dayside around 1815 UT (Figure 2a). The front of the TOI arrives over Svalbard around 1930 UT.

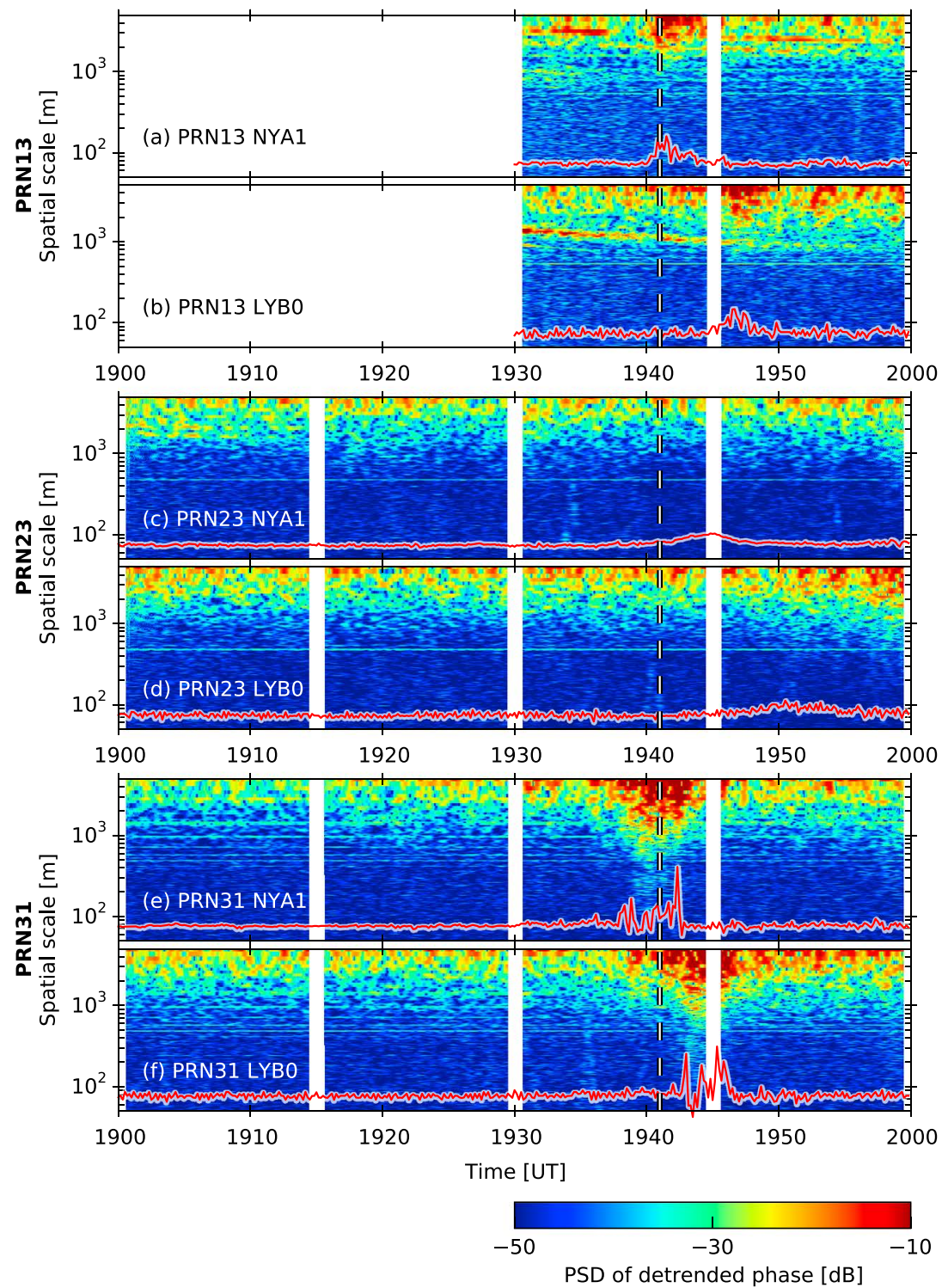

**Figure 5.** Power spectral densities (PSD) of detrended raw 50 Hz phase data for each satellite-receiver pair. The dashed line indicates 1941 UT. The red line shows the 10 s average of 1 Hz ROT (arbitrary units) as an indicator of the TOI leading edge. The white areas in the spectrogram are due to data gaps.

### 4.2. GPS Parameters

Figure 3 presents ASI data showing the propagation of the TOI front and the ionospheric piercing points (IPPs) of the ESR and the GPS satellites at 320 km altitude. The IPP drift speeds at 1941 UT are 167 m/s, 66 m/s, and 60 m/s for PRN13, PRN23, and PRN31, respectively. The drift speed of the plasma is estimated to be 430–480 m/s, based on calculations of the delay between the appearance of the TOI front in the two ESR antennae. Combined with a visual estimation of the plasma drift direction (135–139° geographic east) this yields relative drift velocities of 500–560 m/s, 450–500 m/s, and 460–520 m/s for PRN13, PRN23, and PRN31, respectively.

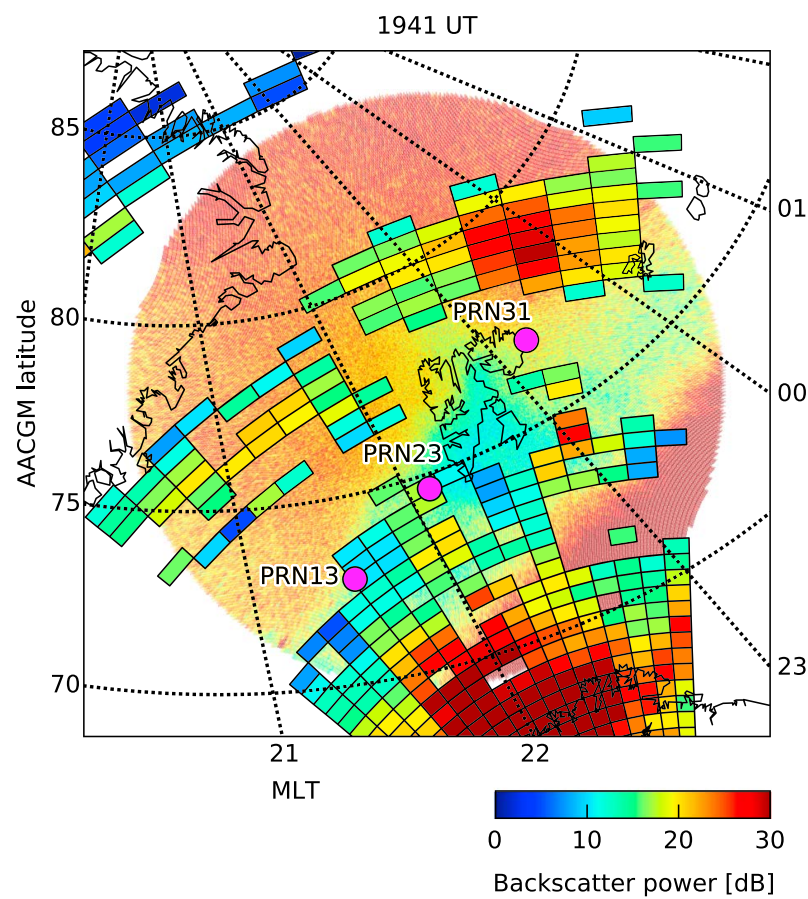

**Figure 6.** SuperDARN backscatter power overlaid on top of all-sky imager data (for reference). GPS ionospheric piercing points are shown together with an Altitude Adjusted Corrected Geomagnetic Coordinates (AACGM) latitude and magnetic local time (MLT) grid.

Figure 4 shows data from the three GPS satellites and the ESR electron density at 320 km altitude. The ESR data (Figures 4a–4c are all identical) show a sudden increase of the electron density to around 4 times the background level as the TOI front passes overhead. In the PRN13 data, we can see an abrupt increase in the VTEC (Figure 4d), also clearly visible in the ROT (Figure 4g), just as the TOI front hits. This is seen first in the two Ny-Ålesund receivers and some minutes later in the Longyearbyen receiver, as one would expect with the current geometry of receivers and drifting plasma structures. Coincident with this TEC gradient at both receiver locations is a burst of weak phase scintillation (Figure 4m), reaching values of $\sigma_\phi = 0.15$ radians before returning to quiet levels just after the TOI front passes and the VTEC stabilizes (and the ROT returns to zero). The $S_4$ amplitude scintillation index (Figure 4j) is practically flat. There are no clear changes in $S_4$ as the TOI front intersects the signal paths.

In the PRN23 data, the VTEC increase (Figure 4e) is softer than in the PRN13 data. This is also seen in the weaker and more gentle ROT increases. There is no indication of any phase or amplitude scintillation (Figures 4k and 4n).

In the PRN31 data, even though the TEC increase (Figure 4f) seems softer than in the PRN13 data, the ROT indicates a structured and steep TEC gradient. Fine structure is also visible in the 1 Hz TEC data from LYB0 and NYA1. There are clear and moderate spikes in $\sigma_\phi$ (Figure 4o) coinciding with the arrival of the TOI front, reaching values of $\sigma_\phi = 0.25$ radians. $S_4$ is very small ($S_4 < 0.1$), though very slight enhancements can be seen colocated with the gradient.

For all satellite-receiver pairs, Figures 4p–4r clearly demonstrates the sensitivity of $\sigma_\phi$ to the detrending cut-off frequency. When detrended with a cutoff frequency of 0.3 Hz, the $\sigma_\phi$ values are reduced significantly compared to detrending using 0.1 Hz.

Figure 5 shows spectrograms of phase variations calculated from detrended raw 50 Hz phase data. The phase variations have been converted to spatial scale sizes by taking into account the relative motion between the plasma and the GPS IPPs. The spectrograms provide a more complete picture of phase variations than $\sigma_\phi$ and are not distorted by the choice of detrending cutoff frequency, since the cutoff frequency only specifies the frequency limit below which (i.e., the spatial scale above which) information is suppressed. A 32,768 m wide Hamming window was used in the calculation of the spectrogram.

In the PRN13 data (Figures 5a and 5b), there is increased power in phase variations at frequencies corresponding to spatial scales of 3–5 km colocated with the gradient (the red lines show the ROT). The PRN23 spectrograms (Figures 5c and 5d) show no such increase in phase variations. The PRN31 spectrograms (Figures 5e and 5f) clearly shows increased power at a wide range of spatial scales, most strongly at ∼5 km but still visible all the way down to ∼100 m, where it reaches the noise level. The SuperDARN data in Figure 6 show that there is strongly enhanced backscatter power (∼30 dB) in the enhanced-density region upstream of PRN31. This is a clear indication of decameter-scale irregularities and supports the wide range of scale sizes shown in the spectrogram.

## 5. Discussion

We have identified the front of a TOI arriving over Svalbard around 1941 UT on 31 October 2011. Electron density data from the ESR show an increase to about 4 times the background density. The signals of three GPS satellites intersected the TOI front almost simultaneously. Phase scintillations were detected at the two satellites experiencing the steepest TEC gradient (PRN13 and PRN31), and the scintillation seemed very well colocated with the TOI leading gradient. The phase spectra showed phase variations at spatial scales of 3–5 km for PRN13 and 100 m–5 km for PRN31.

### 5.1. Cutoff Frequency and $\sigma_\phi$

The increases in $\sigma_\phi$ alone might indicate a structured TOI gradient, but the $\sigma_\phi$ index as it is normally computed is not necessarily a reliable indicator of diffractive structuring at high latitudes [*Forte and Radicella*, 2002; *Forte*, 2005; *Beach*, 2006]. While the $S_4$ amplitude scintillation index is mainly sensitive to irregularity scale sizes at and below the first Fresnel radius, the $\sigma_\phi$ phase scintillation index is highly sensitive to irregularity scale sizes above the Fresnel radius [e.g., *Basu et al.*, 1991]. Such larger-scale sizes correspond to lower-frequency phase fluctuations not pertaining properly to scintillations [e.g., *Forte and Radicella*, 2002]. Thus, the $\sigma_\phi$ index is highly sensitive to the cutoff frequency of the detrending filter, whose function is to remove low-frequency phase variations. The present study clearly exemplifies the dependency of $\sigma_\phi$ on the cutoff frequency (Figures 4m–4r). Examples of the different sensitivity of $S_4$ and $\sigma_\phi$ to the cutoff frequency can be seen in *Forte* [2005].

Several past studies have shown that a detrending cutoff frequency of 0.1 Hz is too low to properly remove refractive phase shifts at high latitudes, which can lead to TEC gradients themselves significantly influencing $\sigma_\phi$ [*Forte and Radicella*, 2002; *Béniguel et al.*, 2004; *Forte*, 2005; *Beach*, 2006]. A cutoff frequency of 0.1 Hz is prevalent in the GNSS community and was originally used in the Wideband satellite experiment of *Fremouw et al.* [1978]. A higher cutoff frequency can mitigate the refractive effects [*Béniguel et al.*, 2009; *Mushini et al.*, 2012]. Other phase scintillation indices based on entirely different calculations or detrending methods have been proposed and may be more suited for high-latitude scintillation studies [*Forte*, 2005; *Mushini et al.*, 2012], but these have not yet been widely used in literature and may have problems of their own [*Mushini et al.*, 2012].

We suggest that spectrograms of detrended phase like Figure 5 may provide a potentially better way to study ionospheric irregularities than phase scintillation indices. First, this avoids the problematic sensitivity of $\sigma_\phi$ to the detrending filter cutoff frequency. Second, spectrograms like Figure 5 provide a more complete view of the irregularity scale sizes present in the plasma, instead of presenting an aggregate number indicating the degree of structuring across a wide range of scale sizes.

### 5.2. Geometrical Effects on Scintillation

Figure 4 showed significant differences in scintillation levels at three different locations (cf. Figure 3). Can geometrical effects cause these differences? While satellites in polar orbits may experience enhanced scintillation from field-aligned irregularities when the signal path aligns with the magnetic field, this has been found not to be a concern for GPS satellites [*Forte and Radicella*, 2004]. GPS scintillation values may however be overestimated at low elevation angles, and while $S_4$ and $\sigma_\phi$ can be projected to the vertical, this is not trivial and depends on, e.g., the anisotropy of the irregularities [*Spogli et al.*, 2009; *Alfonsi et al.*, 2011a]. When the TOI gradient intersected the signal paths, the elevation angles were 25–30°, 46–53°, and 41–42° for PRN13, PRN23, and PRN31, respectively. While PRN23 has the highest elevation angle and the lowest scintillation levels, PRN31 shows higher scintillation levels than PRN13 even though the elevation angle is greater for PRN31 than for PRN13. Thus, the different elevation angles cannot be solely responsible for the observed differences in scintillation levels. Figure 5 indicates that there is in fact structuring at various scales, and we conclude that even though the $\sigma_\phi$ may be overestimated due to detrending or geometrical effects, the variability in $\sigma_\phi$ along the leading edge of the TOI must be real.

### 5.3. Localized Structuring and Plasma Instabilities

The generally low scintillation values (both $\sigma_\phi$ and $S_4$) indicate that the leading TOI gradient is not severely structured. It is however clear from Figure 5 that the front of the TOI is structured at a variety of scale sizes. It is interesting to note the spatial variation of the structuring: First, only the TOI gradient itself seem structured, with no structuring inside the TOI (at least 630–720 km from the front based on PRN31 at 2000 UT,

Figure 5e). Second, the structuring varies significantly along the TOI gradient at distances of several hundred kilometers (the IPP distances are 360 km for PRN13–PRN23 and 550 km for PRN23–PRN31).

It is clear from Figure 5 that the structuring is highly localized to the gradient itself (as indicated by the red line), and that the spectra in the high-density region after the TOI gradient has passed are similar to that of the low-density region before the arrival of the TOI front. While the GDI works on all density gradients, except where the gradient is antiparallel to the plasma flow, it is most effective at trailing edges where the gradient is parallel with the flow. In the case of patches, the GDI may structure the whole patch starting at the trailing edge and working its way toward the leading edge [*Gondarenko and Guzdar*, 2004b]. The present observations indicate that the GDI has not yet developed irregularities from a trailing edge or a structured area through the leading edge of the TOI. Instead, the GDI may be working to some degree along the leading edge of the TOI. Since the GDI is not very effective at leading edges of drifting plasma structures, this means that the GDI may be enhanced by mesoscale flows not visible in our data [e.g., *Grocott et al.*, 2004; *Lyons et al.*, 2011; *Zou et al.*, 2009; *Nishimura et al.*, 2014]. Another possibility is that the leading TOI gradient may not always or everywhere be entirely antiparallel to the plasma flow. Figure 3 shows this: While the leading TOI edge seems straight and perpendicular to the flow in Figure 3c, it is clear from Figures 3a and 3b that this was not the case some minutes earlier. This has implications for the effectiveness of the GDI and could provide an explanation for the differences seen at three different locations along the TOI front. In other words, it may be important to look at the history of the TOI edge morphology.

It is also possible that other instability processes may cause the observed irregularities. *Nishimura et al.* [2014] found flow shears along the edges of a polar cap patch, which persisted as the patch drifted across the polar cap. Such flow shears may be a source of energy for the KHI. *Oksavik et al.* [2011, 2012] have suggested that the KHI may be efficient at spatial scales greater than a few kilometers, so this process may not be important for the structuring we observe at the smallest scales (hundreds of meters). However, our data set cannot exclude the KHI mechanism. In situ measurements [e.g., *Moen et al.*, 2012] of the TOI would be desirable.

### 5.4. HF Backscatter Versus Amplitude Scintillation

Figure 6 shows HF backscatter in the region close to PRN31. HF signals backscatter from field-aligned decameter-scale irregularities [*Greenwald et al.*, 1995]. Figure 6 then indicates decameter-scale irregularities in the region of the TOI intersecting the PRN31 IPP. Amplitude scintillations arise from irregularities of scale sizes between decameters and hundreds of meters. The PRN31 data (Figure 4l) show no significant amplitude scintillations inside the TOI and only very weak amplitude scintillations at the TOI gradient. This may suggest that the decameter-scale irregularities in this case are not strong enough to cause significant amplitude scintillation of the GPS signal. In situ measurements would be required to quantify the irregularities at these scale sizes.

## 6. Conclusions

We have studied the occurrence and nature of GPS ionospheric scintillation at the leading edge of a tongue of ionization (TOI) on 31 October 2011 over Svalbard. Our results indicate the following:

1. The front of the TOI is structured on scale sizes from tens of meters to several kilometers, and the structuring varies significantly along the leading TOI edge at distances of several hundred kilometers.
2. The scintillation and structuring are highly localized to the TOI density gradient. There is no scintillation and no structuring of impact to GPS signals seen well inside the high-density TOI (hundreds of kilometers from the TOI leading edge) as it traverses the nightside polar cap.
3. The phase spectra are similar in the low-density and high-density regions inside and in front of the TOI. It is only at the TOI edge that the spectra differ.
4. We also suggest that spectrograms of detrended phase may be a better way to characterize ionospheric irregularities than is the $\sigma_\phi$ index, which is sensitive to the detrending filter cutoff frequency.

In situ measurements of the TOI, such as sounding rocket measurements, would be needed to determine which mechanism (GDI and/or KHI) dominates the production of irregularities at the leading gradient of the TOI.

# AGU Journal of Geophysical Research: Space Physics

10.1002/2014JA020114

## Acknowledgments
The EISCAT data are available from http://eiscat.se. The UiO ASI data are available at http://tid.uio.no/plasma/aurora. The IMF and solar wind data were provided by the NASA OMNIWeb service (http://omniweb.gsfc.nasa.gov). The AE index was retrieved from the World Data Center for Geomagnetism, Kyoto (http://wdc.kugi.kyoto-u.ac.jp). Figure 2 was made using the tools at the Virginia Tech SuperDARN website (http://vt.superdarn.org/tiki-index.php). The website also provides information for SuperDARN data access. The GPS data can be made available upon request from the author. This study was supported by the Research Council of Norway under contracts 223252/F50, 212014/F50, 208006/F50, 230935/F50, and 230996/F50. EISCAT is an international association supported by research organizations in China (CRIRP), Finland (SA), Japan (NIPR and STEL), Norway (NFR), Sweden (VR), and the United Kingdom (NERC). The Hankasalmi SuperDARN radar is operated by the Radio and Space Physics Plasma Group at the University of Leicester in conjunction with the Finnish Meteorological Institute, Helsinki. We thank PNRA (Italian National Program for Antarctic Researches) and POLARNET-CNR for supporting the ISACCO GNSS network. We would like to thank Kshitija Deshpande at Virginia Tech for fruitful discussions at the 2014 CEDAR workshop and Bjørn Lybekk at the University of Oslo for providing the UiO GPS data.

Alan Rodger thanks Knut Jacobsen and another reviewer for their assistance in evaluating this paper.

## References

Alfonsi, L., L. Spogli, G. De Franceschi, V. Romano, M. Aquino, A. Dodson, and C. N. Mitchell (2011a), Bipolar climatology of GPS ionospheric scintillation at solar minimum, *Radio Sci.*, *46*, RS0D05, doi:10.1029/2010RS004571.

Alfonsi, L., L. Spogli, J. Tong, G. De Franceschi, V. Romano, A. Bourdillon, M. Le Huy, and C. Mitchell (2011b), GPS scintillation and TEC gradients at equatorial latitudes in April 2006, *Adv. Space Res.*, *47*(10), 1750–1757, doi:10.1016/j.asr.2010.04.020.

Anghel, A., A. Astilean, T. Letia, and A. Komjathy (2008), Near real-time monitoring of the ionosphere using dual frequency GPS data in a Kalman filter approach, in *IEEE International Conference on Automation, Quality and Testing, Robotics, 2008. AQTR 2008*, 54–58, doi:10.1109/AQTR.2008.4588793.

Baker, K. B., and S. Wing (1989), A new magnetic coordinate system for conjugate studies at high latitudes, *J. Geophys. Res.*, *94*(A7), 9139–9143, doi:10.1029/JA094iA07p09139.

Basu, S., S. Basu, E. MacKenzie, W. R. Coley, J. R. Sharber, and W. R. Hoegy (1990), Plasma structuring by the gradient drift instability at high latitudes and comparison with velocity shear driven processes, *J. Geophys. Res.*, *95*(A6), 7799–7818, doi:10.1029/JA095iA06p07799.

Basu, S., S. Basu, E. Costa, C. Bryant, C. E. Valladares, and R. C. Livingston (1991), Interplanetary magnetic field control of drifts and anisotropy of high-latitude irregularities, *Radio Sci.*, *26*(4), 1079–1103, doi:10.1029/91RS00586.

Basu, S., S. Basu, P. K. Chaturvedi, and C. M. Bryant (1994), Irregularity structures in the cusp/cleft and polar cap regions, *Radio Sci.*, *29*(1), 195–207, doi:10.1029/93RS01515.

Basu, S., E. J. Weber, T. W. Bullett, M. J. Keskinen, E. MacKenzie, P. Doherty, R. Sheehan, H. Kuenzler, P. Ning, and J. Bongiolatti (1998), Characteristics of plasma structuring in the cusp/cleft region at Svalbard, *Radio Sci.*, *33*(6), 1885–1899, doi:10.1029/98RS01597.

Beach, T. L. (2006), Perils of the GPS phase scintillation index ($\sigma_\phi$), *Radio Sci.*, *41*, RS5S31, doi:10.1029/2005RS003356.

Béniguel, Y., B. Forte, S. M. Radicella, H. J. Strangeways, V. E. Gherm, and N. N. Zernov (2004), Scintillations effects on satellite to earth links for telecommunication and navigation purposes, *Ann. Geophys.*, *47*(2-3), 1179–1199, doi:10.4401/ag-3293.

Béniguel, Y., et al. (2009), Ionospheric scintillation monitoring and modelling, *Ann. Geophys.*, *52*(3-4), 391–416, doi:10.4401/ag-4595.

Briggs, B., and I. Parkin (1963), On the variation of radio star and satellite scintillations with zenith angle, *J. Atmos. Terr. Phys.*, *25*(6), 339–366, doi:10.1016/0021-9169(63)90150-8.

Buchau, J., B. W. Reinisch, E. J. Weber, and J. G. Moore (1983), Structure and dynamics of the winter polar cap f region, *Radio Sci.*, *18*(6), 995–1010, doi:10.1029/RS018i006p00995.

Buchau, J., E. J. Weber, D. N. Anderson, H. C. Carlson, J. G. Moore, B. W. Reinisch, and R. C. Livingston (1985), Ionospheric structures in the polar cap: Their origin and relation to 250-MHz scintillation, *Radio Sci.*, *20*(3), 325–338, doi:10.1029/RS020i003p00325.

Carlson, H. C. (2012), Sharpening our thinking about polar cap ionospheric patch morphology, research, and mitigation techniques, *Radio Sci.*, *47*, RS0L21, doi:10.1029/2011RS004946.

Carlson, H. C., K. Oksavik, J. Moen, A. P. van Eyken, and P. Guio (2002), ESR mapping of polar-cap patches in the dark cusp, *Geophys. Res. Lett.*, *29*(10), 24–1–24–4, doi:10.1029/2001GL014087.

Carlson, H. C., K. Oksavik, J. Moen, and T. Pedersen (2004), Ionospheric patch formation: Direct measurements of the origin of a polar cap patch, *Geophys. Res. Lett.*, *31*, L08806, doi:10.1029/2003GL018166.

Carlson, H. C., J. Moen, K. Oksavik, C. P. Nielsen, I. W. McCrea, T. R. Pedersen, and P. Gallop (2006), Direct observations of injection events of subauroral plasma into the polar cap, *Geophys. Res. Lett.*, *33*, L05103, doi:10.1029/2005GL025230.

Carlson, H. C., T. Pedersen, S. Basu, M. Keskinen, and J. Moen (2007), Case for a new process, not mechanism, for cusp irregularity production, *J. Geophys. Res.*, *112*, A11304, doi:10.1029/2007JA012384.

Carlson, H. C., K. Oksavik, and J. Moen (2008), On a new process for cusp irregularity production, *Ann. Geophys.*, *26*(9), 2871–2885, doi:10.5194/angeo-26-2871-2008.

Carrano, C. S., and K. Groves (2006), The GPS segment of the AFRL-SCINDA global network and the challenges of real-time TEC estimation in the equatorial ionosphere, *Proceedings of the 2006 National Technical Meeting of The Institute of Navigation*, pp. 1036–1047, Monterey, Calif.

Carrano, C. S., A. Anghel, R. A. Quinn, and K. M. Groves (2009), Kalman filter estimation of plasmaspheric total electron content using GPS, *Radio Sci.*, *44*, RS0A10, doi:10.1029/2008RS004070.

Chisham, G., et al. (2007), A decade of the super dual auroral radar network (SuperDARN): Scientific achievements, new techniques and future directions, *Surv. Geophys.*, *28*(1), 33–109, doi:10.1007/s10712-007-9017-8.

Coker, C., G. S. Bust, R. A. Doe, and T. L. Gaussiran (2004), High-latitude plasma structure and scintillation, *Radio Sci.*, *39*, RS1S15, doi:10.1029/2002RS002833.

Cousins, E. D. P., and S. G. Shepherd (2010), A dynamical model of high-latitude convection derived from SuperDARN plasma drift measurements, *J. Geophys. Res.*, *115*, A12329, doi:10.1029/2010JA016017.

De Franceschi, G., L. Alfonsi, V. Romano, M. Aquino, A. Dodson, C. N. Mitchell, P. Spencer, and A. W. Wernik (2008), Dynamics of high-latitude patches and associated small-scale irregularities during the October and November 2003 storms, *J. Atmos. Sol. Terr. Phys.*, *70*(6), 879–888, doi:10.1016/j.jastp.2007.05.018.

Dungey, J. W. (1961), Interplanetary magnetic field and the auroral zones, *Phys. Rev. Lett.*, *6*(2), 47–48, doi:10.1103/PhysRevLett.6.47.

Forte, B. (2005), Optimum detrending of raw GPS data for scintillation measurements at auroral latitudes, *J. Atmos. Sol. Terr. Phys.*, *67*(12), 1100–1109, doi:10.1016/j.jastp.2005.01.011.

Forte, B., and S. M. Radicella (2002), Problems in data treatment for ionospheric scintillation measurements, *Radio Sci.*, *37*(6), 1096, doi:10.1029/2001RS002508.

Forte, B., and S. M. Radicella (2004), Geometrical control of scintillation indices: What happens for GPS satellites, *Radio Sci.*, *39*, RS5014, doi:10.1029/2002RS002852.

Foster, J. C. (1993), Storm time plasma transport at middle and high latitudes, *J. Geophys. Res.*, *98*(A2), 1675–1689, doi:10.1029/92JA02032.

Foster, J. C., and J. R. Doupnik (1984), Plasma convection in the vicinity of the dayside cleft, *J. Geophys. Res.*, *89*(A10), 9107–9113, doi:10.1029/JA089iA10p09107.

Foster, J. C., et al. (2005), Multiradar observations of the polar tongue of ionization, *J. Geophys. Res.*, *110*, A09S31, doi:10.1029/2004JA010928.

Franceschi, G. D., L. Alfonsi, and V. Romano (2006), ISACCO: An Italian project to monitor the high latitudes ionosphere by means of GPS receivers, *GPS Solutions*, *10*(4), 263–267, doi:10.1007/s10291-006-0036-6.

Fremouw, E. J., R. L. Leadabrand, R. C. Livingston, M. D. Cousins, C. L. Rino, B. C. Fair, and R. A. Long (1978), Early results from the DNA wideband satellite experiment-Complex-signal scintillation, *Radio Sci.*, *13*(1), 167–187, doi:10.1029/RS013i001p00167.

Gondarenko, N. A., and P. N. Guzdar (2004a), Density and electric field fluctuations associated with the gradient drift instability in the high-latitude ionosphere, *Geophys. Res. Lett.*, *31*, L11802, doi:10.1029/2004GL019703.

Gondarenko, N. A., and P. N. Guzdar (2004b), Plasma patch structuring by the nonlinear evolution of the gradient drift instability in the high-latitude ionosphere, *J. Geophys. Res.*, *109*, A09301, doi:10.1029/2004JA010504.

Greenwald, R. A., et al. (1995), DARN/SuperDARN, *Space Sci. Rev.*, *71*(1-4), 761–796, doi:10.1007/BF00751350.

Grocott, A., T. K. Yeoman, R. Nakamura, S. W. H. Cowley, H. U. Frey, H. Rème, and B. Klecker (2004), Multi-instrument observations of the ionospheric counterpart of a bursty bulk flow in the near-earth plasma sheet, *Ann. Geophys.*, *22*(4), 1061–1075, doi:10.5194/angeo-22-1061-2004.

Hey, J. S., S. J. Parsons, and J. W. Phillips (1946), Fluctuations in cosmic radiation at radio-frequencies, *Nature*, *158*(4007), 234–234, doi:10.1038/158234a0.

Hosokawa, K., K. Shiokawa, Y. Otsuka, A. Nakajima, T. Ogawa, and J. D. Kelly (2006), Estimating drift velocity of polar cap patches with all-sky airglow imager at Resolute Bay, Canada, *Geophys. Res. Lett.*, *33*, L15111, doi:10.1029/2006GL026916.

Hosokawa, K., K. Shiokawa, Y. Otsuka, T. Ogawa, J.-P. St-Maurice, G. J. Sofko, and D. A. Andre (2009), Relationship between polar cap patches and field-aligned irregularities as observed with an all-sky airglow imager at Resolute Bay and the PolarDARN radar at Rankin inlet, *J. Geophys. Res.*, *114*, A03306, doi:10.1029/2008JA013707.

Hosokawa, K., J. I. Moen, K. Shiokawa, and Y. Otsuka (2011), Decay of polar cap patch, *J. Geophys. Res.*, *116*, A05306, doi:10.1029/2010JA016297.

Keskinen, M. J., and S. L. Ossakow (1983), Theories of high-latitude ionospheric irregularities: A review, *Radio Sci.*, *18*(6), 1077–1091, doi:10.1029/RS018i006p01077.

Kintner, P. M., B. M. Ledvina, and E. R. de Paula (2007), GPS and ionospheric scintillations, *Space Weather*, *5*, S09003, doi:10.1029/2006SW000260.

Knudsen, W. C. (1974), Magnetospheric convection and the high-latitude F2 ionosphere, *J. Geophys. Res.*, *79*(7), 1046–1055, doi:10.1029/JA079i007p01046.

Lehtinen, M. S., and A. Huuskonen (1996), General incoherent scatter analysis and GUISDAP, *J. Atmos. Terr. Phys.*, *58*(1-4), 435–452, doi:10.1016/0021-9169(95)00047-X.

Lepping, R. P., et al. (1995), The WIND magnetic field investigation, *Space Sci. Rev.*, *71*(1-4), 207–229, doi:10.1007/BF00751330.

Li, G., B. Ning, Z. Ren, and L. Hu (2010), Statistics of GPS ionospheric scintillation and irregularities over polar regions at solar minimum, *GPS Solutions*, *14*(4), 331–341, doi:10.1007/s10291-009-0156-x.

Lockwood, M., and H. C. Carlson (1992), Production of polar cap electron density patches by transient magnetopause reconnection, *Geophys. Res. Lett.*, *19*(17), 1731–1734, doi:10.1029/92GL01993.

Lockwood, M., J. A. Davies, J. Moen, A. P. van Eyken, K. Oksavik, I. W. McCrea, and M. Lester (2005), Motion of the dayside polar cap boundary during substorm cycles: II. Generation of poleward-moving events and polar cap patches by pulses in the magnetopause reconnection rate, *Ann. Geophys.*, *23*(11), 3513–3532, doi:10.5194/angeo-23-3513-2005.

Lorentzen, D. A., N. Shumilov, and J. Moen (2004), Drifting airglow patches in relation to tail reconnection, *Geophys. Res. Lett.*, *31*, L02806, doi:10.1029/2003GL017785.

Lorentzen, D. A., J. Moen, K. Oksavik, F. Sigernes, Y. Saito, and M. G. Johnsen (2010), In situ measurement of a newly created polar cap patch, *J. Geophys. Res.*, *115*, A12323, doi:10.1029/2010JA015710.

Lyons, L. R., Y. Nishimura, H.-J. Kim, E. Donovan, V. Angelopoulos, G. Sofko, M. Nicolls, C. Heinselman, J. M. Ruohoniemi, and N. Nishitani (2011), Possible connection of polar cap flows to pre- and post-substorm onset PBIs and streamers, *J. Geophys. Res.*, *116*, A12225, doi:10.1029/2011JA016850.

Millward, G. H., R. J. Moffett, H. F. Balmforth, and A. S. Rodger (1999), Modeling the ionospheric effects of ion and electron precipitation in the cusp, *J. Geophys. Res.*, *104*(A11), 24,603–24,612, doi:10.1029/1999JA900249.

Mitchell, C. N., L. Alfonsi, G. De Franceschi, M. Lester, V. Romano, and A. W. Wernik (2005), GPS TEC and scintillation measurements from the polar ionosphere during the October 2003 storm, *Geophys. Res. Lett.*, *32*, L12S03, doi:10.1029/2004GL021644.

Moen, J., H. C. Carlson, K. Oksavik, C. P. Nielsen, S. E. Pryse, H. R. Middleton, I. W. McCrea, and P. Gallop (2006), EISCAT observations of plasma patches at sub-auroral cusp latitudes, *Ann. Geophys.*, *24*(9), 2363–2374, doi:10.5194/angeo-24-2363-2006.

Moen, J., N. Gulbrandsen, D. A. Lorentzen, and H. C. Carlson (2007), On the MLT distribution of f region polar cap patches at night, *Geophys. Res. Lett.*, *34*, L14113, doi:10.1029/2007GL029632.

Moen, J., X. C. Qiu, H. C. Carlson, R. Fujii, and I. W. McCrea (2008), On the diurnal variability in F2-region plasma density above the EISCAT Svalbard radar, *Ann. Geophys.*, *26*(8), 2427–2433, doi:10.5194/angeo-26-2427-2008.

Moen, J., K. Oksavik, T. Abe, M. Lester, Y. Saito, T. A. Bekkeng, and K. S. Jacobsen (2012), First in-situ measurements of HF radar echoing targets, *Geophys. Res. Lett.*, *39*, L07104, doi:10.1029/2012GL051407.

Moen, J., K. Oksavik, L. Alfonsi, Y. Daabakk, V. Romano, and L. Spogli (2013), Space weather challenges of the polar cap ionosphere, *J. Space Weather Space Clim.*, *3*, A02, doi:10.1051/swsc/2013025.

Mushini, S. C., P. T. Jayachandran, R. B. Langley, J. W. MacDougall, and D. Pokhotelov (2012), Improved amplitude- and phase-scintillation indices derived from wavelet detrended high-latitude GPS data, *GPS Solutions*, *16*(3), 363–373, doi:10.1007/s10291-011-0238-4.

Nishimura, Y., et al. (2014), Day-night coupling by a localized flow channel visualized by polar cap patch propagation, *Geophys. Res. Lett.*, *41*, 3701–3709, doi:10.1002/2014GL060301.

NovAtel (2007), *GSV4004B GPS Ionospheric Scintillation & TEC Monitor User's Manual*.

Ogilvie, K. W., et al. (1995), SWE, a comprehensive plasma instrument for the WIND spacecraft, *Space Sci. Rev.*, *71*(1-4), 55–77, doi:10.1007/BF00751326.

Oksavik, K., V. L. Barth, J. Moen, and M. Lester (2010), On the entry and transit of high-density plasma across the polar cap, *J. Geophys. Res.*, *115*, A12308, doi:10.1029/2010JA015817.

Oksavik, K., J. I. Moen, E. H. Rekaa, H. C. Carlson, and M. Lester (2011), Reversed flow events in the cusp ionosphere detected by SuperDARN HF radars, *J. Geophys. Res.*, *116*, A12303, doi:10.1029/2011JA016788.

Oksavik, K., J. Moen, M. Lester, T. A. Bekkeng, and J. K. Bekkeng (2012), In situ measurements of plasma irregularity growth in the cusp ionosphere, *J. Geophys. Res.*, *117*, A11301, doi:10.1029/2012JA017835.

Ossakow, S. L., and P. K. Chaturvedi (1979), Current convective instability in the diffuse aurora, *Geophys. Res. Lett.*, *6*(4), 332–334, doi:10.1029/GL006i004p00332.

Prikryl, P., P. T. Jayachandran, S. C. Mushini, D. Pokhotelov, J. W. MacDougall, E. Donovan, E. Spanswick, and J.-P. St.-Maurice (2010), GPS TEC, scintillation and cycle slips observed at high latitudes during solar minimum, *Ann. Geophys.*, *28*(6), 1307–1316, doi:10.5194/angeo-28-1307-2010.

Prikryl, P., P. T. Jayachandran, S. C. Mushini, and R. Chadwick (2011a), Climatology of GPS phase scintillation and HF radar backscatter for the high-latitude ionosphere under solar minimum conditions, *Ann. Geophys.*, *29*(2), 377–392, doi:10.5194/angeo-29-377-2011.

Prikryl, P., et al. (2011b), Interhemispheric comparison of GPS phase scintillation at high latitudes during the magnetic-cloud-induced geomagnetic storm of 5-7 April 2010, *Ann. Geophys.*, *29*(12), 2287–2304, doi:10.5194/angeo-29-2287-2011.

Romano, V., S. Pau, M. Pezzopane, E. Zuccheretti, B. Zolesi, G. De Franceschi, and S. Locatelli (2008), The electronic space weather upper atmosphere (eSWua) project at INGV: Advancements and state of the art, *Ann. Geophys.*, *26*(2), 345–351, doi:10.5194/angeo-26-345-2008.

Romano, V., S. Pau, M. Pezzopane, L. Spogli, E. Zuccheretti, M. Aquino, and C. M. Hancock (2013), eSWua: A tool to manage and access GNSS ionospheric data from mid-to-high latitudes, *Ann. Geophys.*, *56*(2), R0223, doi:10.4401/ag-6244.

Sato, T. (1959), Morphology of the ionospheric f2 disturbances in the polar region, *Rep. Ionos. Space Res. Jpn.*, *13*, 91–95.

Spogli, L., L. Alfonsi, G. De Franceschi, V. Romano, M. H. O. Aquino, and A. Dodson (2009), Climatology of GPS ionospheric scintillations over high and mid-latitude European regions, *Ann. Geophys.*, *27*(9), 3429–3437, doi:10.5194/angeo-27-3429-2009.

Tiwari, S., A. Jain, S. Sarkar, S. Jain, and A. K. Gwal (2012), Ionospheric irregularities at antarctic using GPS measurements, *J. Earth Syst. Sci.*, *121*(2), 345–353, doi:10.1007/s12040-012-0168-8.

Tsunoda, R. T. (1988), High-latitude F region irregularities: A review and synthesis, *Rev. Geophys.*, *26*(4), 719–760, doi:10.1029/RG026i004p00719.

Weber, E. J., J. Buchau, J. G. Moore, J. R. Sharber, R. C. Livingston, J. D. Winningham, and B. W. Reinisch (1984), F layer ionization patches in the polar cap, *J. Geophys. Res.*, *89*(A3), 1683–1694, doi:10.1029/JA089iA03p01683.

Weber, E. J., J. A. Klobuchar, J. Buchau, H. C. Carlson, R. C. Livingston, O. de la Beaujardiere, M. McCready, J. G. Moore, and G. J. Bishop (1986), Polar cap f layer patches: Structure and dynamics, *J. Geophys. Res.*, *91*(A11), 12,121–12,129, doi:10.1029/JA091iA11p12121.

Wickwar, V. B., L. L. Cogger, and H. C. Carlson (1974), The 6300 å O1D airglow and dissociative recombination, *Planet. Space Sci.*, *22*(5), 709–724, doi:10.1016/0032-0633(74)90141-X.

Zhang, Q.-H., et al. (2013a), Direct observations of the evolution of polar cap ionization patches, *Science*, *339*(6127), 1597–1600, doi:10.1126/science.1231487, PMID: 23539601.

Zhang, Q.-H., B.-C. Zhang, J. Moen, M. Lockwood, I. W. McCrea, H.-G. Yang, H.-Q. Hu, R.-Y. Liu, S.-R. Zhang, and M. Lester (2013b), Polar cap patch segmentation of the tongue of ionization in the morning convection cell, *Geophys. Res. Lett.*, *40*, 2918–2922, doi:10.1002/grl.50616.

Zou, S., L. R. Lyons, M. J. Nicolls, and C. J. Heinselman (2009), PFISR observations of strong azimuthal flow bursts in the ionosphere and their relation to nightside aurora, *J. Atmos. Sol. Terr. Phys.*, *71*(6-7), 729–737, doi:10.1016/j.jastp.2008.06.015.